\documentclass[aps,11pt,preprintnumbers,nofootinbib,floatfix]{revtex4}

\usepackage{graphicx}% Include figure files
\usepackage{epsfig}
\usepackage{dcolumn}% Align table columns on decimal point
\usepackage{subfigure}
\usepackage{multirow}
\usepackage{amsmath}
\usepackage{amssymb}

\begin{document}

\title{Effects of massive gravity on $p$-wave holographic superconductor}

%%%% To generate auto affiliation numbers please use \author{}\affil{} command

\author{Cao H. Nam}
\email{nam.caohoang@phenikaa-uni.edu.vn}  
\affiliation{Phenikaa Institute for Advanced Study and Faculty of Fundamental Sciences, Phenikaa University, Yen Nghia, Ha Dong, Hanoi 12116, Vietnam}
\date{\today}

\begin{abstract}%
In this paper, we have analytically investigated $p$-wave holographic superconductor in the framework of massive gravity in the probe limit. We obtained the analytical expressions for the critical temperature, the value of the condensate operator, and the difference of the free energy between the superconductor and normal phases. We studied the behavior of these quantities in the presence of the mass of graviton, which is found to be dependent almost on the sign of the massive gravity couplings. The critical temperature becomes (lower)higher and the condensate value gets (larger)smaller with (decreasing)increasing the massive gravity couplings or increasing the mass of graviton with the sufficiently (negative)positive massive gravity couplings. This fact corresponds to that the superconductor phase is (more)less thermodynamically favored.
\end{abstract}

\maketitle

\section{Introduction}
Massive gravity is a modification of Einstein gravity in the infrared (IR) region with including a mass to graviton. As a result, this modification leads to new physical degrees of freedom and thus it has profound consequences such as the natural resolution for the acceleration of our universe without introducing dark energy. The recently direct detections of the gravitational waves by LIGO on the binary black hole merger have provided an upper bound on the mass of graviton as, $m_g\leq1.2\times10^{-22}$ eV \cite{LIGO}. This has made the question regarding whether graviton has mass become more interesting. The first attempt to construct massive gravity was done by Fierz and Pauli \cite{Fierz1939}. Unfortunately, this massive gravity suffers from a pathology well-known as the van Dam-Veltman-Zakharov (vDVZ) discontinuity \cite{Dam1970,Zakharov1970}.
As showed by Vainshtein, the vDVZ discontinuity could be resolved in a nonlinear framework by introducing higher-order interaction terms \cite{Vainshtein1972}. However, this nonlinear massive gravity encountered the Boulware-Deser (BD) ghost \cite{Boulware1972}. In $2010$, de Rham, Gabadadze and Tolley proposed a successful nonlinear massive gravity theory which avoids the vDVZ discontinuity and the BD ghost \cite{deRham2010,deRham2011}.

The AdS/CFT correspondence \cite{Maldacena} determines a relationship between the weakly coupled gravity theory in $(d+1)$-dimensional bulk AdS spacetime and the strongly coupled conformal field theory living on the $d$-dimensional boundary. Hence, this correspondence has been regarded as a powerful tool to explore physics of the strongly correlated system by using the weakly interacting gravitational dual in one higher
dimension. Based on the ideal of the AdS/CFT correspondence, Hartnoll \emph{et al.} built a model of $s$-wave holographic superconductor at which a complex scalar field is coupled to the $U(1)$ gauge field in the framework of Einstein gravity \cite{Horowitz2008a,Horowitz2008b}. It was found that below a critical temperature RN-AdS black brane which is dual to the normal phase in the boundary theory is the instability due to the condensate of the scalar field. Black brane with this nontrivial configuration of the scalar field is dual to the superconductor phase in the boundary theory. The model of $s$-wave holographic superconductor has been generalized to investigate $p$-wave holographic superconductor at which the spin-$1$ order parameter in the boundary theory is dual to a $2$-form field \cite{Donos2011}, a $SU(2)$ Yang-Mills gauge field \cite{Pufu2008}, or a complex vector field \cite{CaiLi2013,CaiLi2014} in the bulk theory.

The properties of holographic superconductors are dependent on the background of the black hole geometry as well as the action form of the $U(1)$ gauge field. Since the investigation of holographic superconductors has received much attention in the literature, which are considered in alternative theories of gravity as well as in the nonlinear electrodynamics. The $p$-wave and $s$-wave holographic superconductor models were investigated in Einstein-Gauss-Bonnet gravity \cite{Soda2009,CaiZhang2010,Barclay2010,PanWang2010,PanChen2011a,JingChen2012,CuiXue2013,Ghorai2019,LiuJing2017} and in the nonlinear electrodynamics \cite{PanChen2011b,Banerjee2013,Ghorai2016,Asl2018,Mohammadi2019,Nam2019,Srivastav2020}. Also, $s$-wave holographic superconductor was studied in massive gravity \cite{ZengWu2014,LiZhao2019}, in the background of the Lifshitz black hole geometry \cite{ZhaoJing2014}, and in the background of the black hole surrounded by string cloud \cite{Nam2019b}. In addition, $p$-wave holographic superconductor was considered in the background of the scalar hairy black hole \cite{WenQian2019}.

Despite the great success of Einstein gravity at the large distances or in the IR region, there is the consideration of alternatives to Einstein gravity in this region, which is motivated by the phenomenological reasons. As mentioned above, one of IR modifications of Einstein gravity is massive gravity which includes the mass terms for graviton. Therefore, it is interesting to generalize the investigation of holographic superconductor to massive gravity. In addition, an interesting aspect of holographic superconductor in massive gravity is that the momentum dissipation effect of a lattice can be incoporated because the presence of the mass of graviton breaks the translation symmetry \cite{Vegh1301}. With these interesting aspects, the investigation of holographic superconductor in massive gravity has got attentions \cite{ZengWu2014,LiZhao2019}, however, only the case of $s$-wave has been explored so far. Hence, in this work we will construct a $p$-wave holographic superconductor model in massive gravity and study the role and physical impact of the graviton mass terms on the properties of $p$-wave holographic superconductor.

This paper is organized as follows. In Sect. \ref{HDM}, we introduce a $p$-wave holographic superconductor model in the probe limit in the black brane geometry background with the presence of the mass of graviton. In Sect. \ref{CT}, we use the Sturm-Liouville analytical method to obtain an expression for the critical temperature as a function of the charge density and then investigate the presence of the mass of graviton on the behavior of the critical temperature. In Sect. \ref{CV}, we determine analytically the condensate value and study its behavior in the presence of the mass of graviton. In Sect. \ref{FE}, we calculate the free energy for the superconductor and normal phases. Finally, we conclude our main results in Sect. \ref{conclu}.

\section{\label{HDM}Holographic dual model}
In this section, we construct a $p$-wave holographic superconductor model which is considered in the context of massive gravity. The corresponding action is described by
\begin{equation}
S=\frac{1}{2}\int
d^dx\sqrt{-g}\left[R+\frac{(d-1)(d-2)}{l^2}+m^2_g\sum^4_{i=1}c_i\mathcal{U}_i(g,f)+\mathcal{L}_{\text{mat}}\right],\label{EMG-nlED-adS}
\end{equation}
where $R$ is the scalar curvature of the spacetime, $l$ is the AdS radius, $m_g$ is the mass of graviton, $c_i$ are the coupling parameters, $f$ is the reference metric which is of course a symmetric tensor, $\mathcal{U}_i$ are symmetric polynomials in terms of the eigenvalues of the $4\times4$ matrix
${\mathcal{K}^\mu}_\nu=\sqrt{g^{\mu\lambda}f_{\lambda\nu}}$ given
as
\begin{eqnarray}
\mathcal{U}_1&=&[\mathcal{K}],\nonumber \\
\mathcal{U}_2&=&[\mathcal{K}]^2-[\mathcal{K}^2],\nonumber \\
\mathcal{U}_3&=&[\mathcal{K}]^3-3[\mathcal{K}][\mathcal{K}^2]+2[\mathcal{K}^3],\nonumber\\
\mathcal{U}_4&=&[\mathcal{K}]^4-6[\mathcal{K}]^2[\mathcal{K}^2]+8[\mathcal{K}][\mathcal{K}^3]+3[\mathcal{K}^2]^2-6[\mathcal{K}^4],\nonumber\label{masgrav-pols}
\end{eqnarray}
with $[\mathcal{K}]={\mathcal{K}^\mu}_\mu$. The matter term is given as
\begin{equation}
\mathcal{L}_{\text{mat}}=-\frac{1}{4}F_{\mu\nu}F^{\mu\nu}-\frac{1}{2}\left(D_\mu\rho_\nu-D_\nu\rho_\mu\right)^\dagger\left(D^\mu\rho^\nu-D^\nu\rho^\mu\right)-m^2\rho^\dagger_\mu\rho^\mu+iq\gamma\rho_\mu\rho^\dagger_\nu F^{\mu\nu},\label{matt-Lag}
\end{equation}
where $F_{\mu\nu}=\nabla_\mu A_\nu-\nabla_\nu A_\mu$ with $A_\mu$ representing the $U(1)$ gauge field, $\rho_\mu$ is the complex vector field of the mass $m$ and the charge $q$, $D_\mu=\nabla_\mu-iqA_\mu$, and $\gamma$ refers to the magnetic moment of the complex vector field $\rho_\mu$. In this work we consider the case without the external magnetic field and hence the last term in (\ref{matt-Lag}) is set to zero.

We find a black brane solution with the metric ansatz
\begin{eqnarray}
ds^2=-f(r)dt^2+\frac{dr^2}{f(r)}+r^2h_{ij}dx^idx^j,
\end{eqnarray}
where $h_{ij}dx^idx^j=dx^2_1+dx^2_2+...+dx^2_{d-2}$ is the line element of the $(d-2)$-dimensional planar hypersurface. Following Ref. \cite{Vegh1301}, we consider the reference metric in the form
as $f_{\mu\nu}=\text{diag}(0,0,h_{ij})$. In addition, we adopt the following ansatz for the gauge field and complex vector field
\begin{eqnarray}
A_\mu=\phi(r)\delta^t_\mu,\ \ \ \ \rho_\mu=\rho_x(r)\delta^x_\mu.
\end{eqnarray}

By varying the action (\ref{EMG-nlED-adS}) with the ansatz for the fields given above, we obtain the equations of motion in the probe limit as
\begin{eqnarray}
&&\frac{d-2}{2r^4}\left[r^2(d-3)f+r^3f'\right]-\frac{(d-1)(d-2)}{2l^2}-\frac{m^2_g}{2}\left[\frac{(d-2)c_1}{r}+\frac{(d-2)(d-3)c_2}{r^2}\right.\nonumber\\
&&\left.\frac{(d-2)(d-3)(d-4)c_3}{r^3}+\frac{(d-2)(d-3)(d-4)(d-5)c_4}{r^4}\right]=0,\label{Enseq}\\
%&&\left(1+3b\phi'^2\right)\phi''+\frac{d-2}{r}\left(1+b\phi'^2\right)\phi'-\frac{2q^2\rho^2_x}{r^2f}\phi=0,%\label{phi-eq}\\
&&\phi''(r)+\frac{d-2}{r}\phi'(r)-\frac{2q^2\rho^2_x(r)}{r^2f(r)}\phi(r)=0,\label{phi-eq}\\
&&\rho''_x(r)+\left(\frac{f'(r)}{f(r)}+\frac{d-4}{r}\right)\rho'_x(r)+\left(\frac{q^2\phi^2(r)}{f^2(r)}-\frac{m^2}{f(r)}\right)\rho_x(r)=0.\label{rhox-eq}
\end{eqnarray}
Eq. (\ref{Enseq}) leads to the following metric function $f(r)$ \cite{CaiHu2013}
\begin{eqnarray}
f(r)&=&\frac{r^2}{l^2}\left(1-\frac{r^{d-1}_+}{r^{d-1}}\right)+m^2_g\left[\frac{c_1r}{d-2}\left(1-\frac{r^{d-2}_+}{r^{d-2}}\right)+c_2\left(1-\frac{r^{d-3}_+}{r^{d-3}}\right)\right.\nonumber\\
&&\left.+\frac{(d-3)c_3}{r}\left(1-\frac{r^{d-4}_+}{r^{d-4}}\right)+\frac{(d-3)(d-4)c_4}{r^2}\left(1-\frac{r^{d-5}_+}{r^{d-5}}\right)\right],
\end{eqnarray}
where $r_+$ is the event horizon radius. The temperature of the boundary field theory is identified as the Hawking temperature of the black brane and it is given by
\begin{eqnarray}
T_H=\frac{(d-1)r_+}{4\pi l^2}+\frac{m^2_g}{4\pi}\left[c_1+\frac{(d-3)c_2}{r_+}+\frac{(d-3)(d-4)c_3}{r^2_+}+\frac{(d-3)(d-4)(d-5)c_4}{r^3_+}\right].
\end{eqnarray}

For the fields well-behaving at the event horizon, we require the regularity condition as, $\phi(r_+)=0$ and $\rho'_x(r_+)=m^2\rho_x(r_+)/f'(r_+)$. The asymptotic behavior of the fields near the AdS boundary ($r\rightarrow\infty$) is given by
\begin{eqnarray}
\phi(r)&=&\mu-\frac{\rho}{r^{d-3}},\label{phi-asybeh}\\
\rho_x(r)&=&\frac{\langle\mathcal{O}_{x-}\rangle}{r^{\Delta_-}}+\frac{\langle\mathcal{O}_{x+}\rangle}{r^{\Delta_+}},\label{rhox-asymbeh}
\end{eqnarray}
where $\Delta_\pm=\left[d-3\pm\sqrt{(d-3)^2+4m^2l^2}\right]/2$, 
$\mu$ and $\rho$ are the chemical potential and charge density, respectively. It should be noted that first the mass of the complex vector field satisfies the Breitenlohner-Freedman bound \cite{Breitenlohner1982} as,
$m^2\geq-\frac{(d-3)^2}{4l^2}$. Second, according to AdS/CFT correspondence, either $\langle\mathcal{O}_{x-}\rangle$ or $\langle\mathcal{O}_{x+}\rangle$ is interpreted as the source and the other is interpreted as the vacuum expectation value of the $x$-component of the vector operator in the dual field theory at the AdS boundary. From requiring that the condensate appears spontaneously, we impose the vanishing source $\langle\mathcal{O}_{x-}\rangle=0$.
 
By introducing the new coordinate $z=\frac{r_+}{r}$, one can rewrite Eqs. (\ref{phi-eq}) and (\ref{rhox-eq}) as
%\begin{eqnarray}
%\left(1+\frac{3bz^4}{r^2_+}\phi'^2(z)\right)\phi''(z)+\frac{1}{z}\left[(4-d)+\frac{(8-d)bz^4}{r^2_+}
%\phi'^2(z)\right]\phi'(z)
%\end{eqnarray} 
\begin{eqnarray}
&&\phi''(z)+\frac{4-d}{z}\phi'(z)-\frac{2q^2\rho^2_x(z)}{z^2f(z)}\phi(z)=0,\label{phi-eq-z}\\
&&\rho''_x(z)+\left[\frac{6-d}{z}+\frac{f'(z)}{f(z)}\right]\rho'_x(z)+\frac{r^2_+}{z^4}\left[\frac{q^2\phi^2(z)}{f^2(z)}-\frac{m^2}{f(z)}\right]\rho_x(z)=0,\label{rhox-eq-z}
\end{eqnarray} 
where the function $f(z)$ is given by
\begin{eqnarray}
f(z)&=&\frac{r^2_+}{l^2z^2}(1-z^{d-1})+m^2_g\left[\frac{c_1r_+}{(d-2)z}(1-z^{d-2})+c_2(1-z^{d-3})\right.\nonumber\\
&&\left.+\frac{(d-3)c_3z}{r_+}(1-z^{d-4})+\frac{(d-3)(d-4)c_4z^2}{r^2_+}(1-z^{d-5})\right].
\end{eqnarray}
In the following sections, we will solve these equations to investigate the properties of $p$-wave holographic superconductor in the black brane geometry background in the presence of the mass of graviton. In what follows, we fix $q=1$ and $l=1$ which can be chosen by using the scaling symmetries. 

\section{\label{CT}Critical temperature versus charge density}

In this section, we determine the relation between the critical temperature $T_c$, below which the condensate value $\langle\mathcal{O}_{x+}\rangle$ is nonzero, and the charge density $\rho$. In order to obtain this relation, we employ the analytical approach based on the Sturm-Liouville eigenvalue problem.

At the critical temperature $T_c$ the condensate value $\langle\mathcal{O}_{x+}\rangle$ vanishes and hence Eq. (\ref{phi-eq-z}) becomes
\begin{eqnarray}
\phi''(z)+\frac{4-d}{z}\phi'(z)=0.
\end{eqnarray}
The solution of this equation is obtained as
\begin{eqnarray}
\phi(z)=r_{+c}\lambda(1-z^{d-3})\equiv r_{+c}\lambda\xi(z),\label{phi-sol-Tc}
\end{eqnarray}
where $r_{+c}$ is the horizon radius of the black brane with the temperature $T_c$, $\lambda\equiv\frac{\rho}{r^{d-2}_{+c}}$, and we have used the boundary condition $\phi(1)=0$ and the asymptotic behavior given at Eq. (\ref{phi-asybeh}).

In the limit that the temperature is near the critical temperature $T_c$, we can express $\rho_x(z)$ in the following form
\begin{eqnarray}
\rho_x(z)=\langle\mathcal{O}_{x+}\rangle\frac{z^{\Delta_+}}{r^{\Delta_+}_+}F(z),\label{rho-exp}
\end{eqnarray}
where $F(z)$ is a trial function which satisfies the conditions $F(0)=1$ and $F'(0)=0$ consistent to the asymptotic behavior given at Eq. (\ref{rhox-asymbeh}). By substituting this form of $\rho_x(z)$ and the solution of $\phi(z)$ given at Eq. (\ref{phi-sol-Tc}) into Eq. (\ref{rhox-eq-z}), we obtain the equation in the form of Sturm-Liouville equation as
\begin{eqnarray}
\left[T(z)F'(z)\right]'-Q(z)F(z)+\lambda^2P(z)F(z)=0,\label{SL-eq}
\end{eqnarray}
where
\begin{eqnarray}
T(z)&=&\exp\left\{\int dz\left[\frac{6-d+2\Delta_+}{z}+\frac{g'(z)}{g(z)}\right]\right\},\nonumber\\
&=&z^{4-d+2\Delta_+}\left\{1-z^{d-1}+m^2_g\left[\frac{\bar{c}_1z}{d-2}(1-z^{d-2})+\bar{c}_2z^2(1-z^{d-3})\right.\right.\nonumber\\
&&\left.\left.+(d-3)\bar{c}_3z^3(1-z^{d-4})+(d-3)(d-4)\bar{c}_4z^4(1-z^{d-5})\right]\right\},\nonumber\\
Q(z)&=&-T(z)\left[\frac{\Delta_+}{z}\left(\frac{5-d+\Delta_+}{z}+\frac{g'(z)}{g(z)}\right)-\frac{m^2}{z^4g(z)}\right],\nonumber\\
P(z)&=&\frac{T(z)\xi^2(z)}{z^4g^2(z)},\nonumber\\
g(z)&=&\frac{1-z^{d-1}}{z^2}+m^2_g\left[\frac{\bar{c}_1}{(d-2)z}(1-z^{d-2})+\bar{c}_2(1-z^{d-3})\right.\nonumber\\
&&\left.+(d-3)\bar{c}_3z(1-z^{d-4})+(d-3)(d-4)\bar{c}_4z^2(1-z^{d-5})\right],
\end{eqnarray}
with $\bar{c}_n\equiv c_n/r^n_{+c}$ for $n=1,2,3,4$. From the Sturm-Liouville eigenvalue problem, the eigenvalue $\lambda^2$ in Eq. (\ref{SL-eq}) is obtained by minimizing the following expression
\begin{equation}
\lambda^2=\frac{\int^1_0T(z)F'^2(z)dz+\int^1_0Q(z)F^2(z)dz}{\int^1_0P(z)F^2(z)dz},\label{sqlam-func}
\end{equation}
where the trial function $F(z)$ is chosen as $F(z)=1-\beta z^2$ \cite{Therrien2010}. As a result, we can express the critical temperature $T_c$ in terms of the charge density $\rho$ as
\begin{eqnarray}
T_c&=&\frac{1}{4\pi}\left\{d-1+m^2_g\left[\bar{c}_1+(d-3)\bar{c}_2+(d-3)(d-4)\bar{c}_3+(d-3)(d-4)(d-5)\bar{c}_4\right]\right\}\nonumber\\
&&\times\left(\frac{\rho}{\lambda_{\text{min}}}\right)^{\frac{1}{d-2}},\label{critemp}
\end{eqnarray}
where $\lambda_{\text{min}}$ is the minimum value obtained from minimizing Eq. (\ref{sqlam-func}). The effect of massive gravity on the critical temperature $T_c$ is shown through which it enters into both the expression of $T_c$ and $\lambda_{\text{min}}$. Also, it should be noted that only the couplings $c_{1,2}$ affect the critical temperature in the case of four dimensions. Whereas, the couplings $c_3$ and $c_4$ modify the critical temperature if the spacetime dimension is larger than four and five, respectively. 

\begin{figure}[t]
% Requires \usepackage{graphicx}\
 \centering
\begin{tabular}{cc}
\includegraphics[width=0.48 \textwidth]{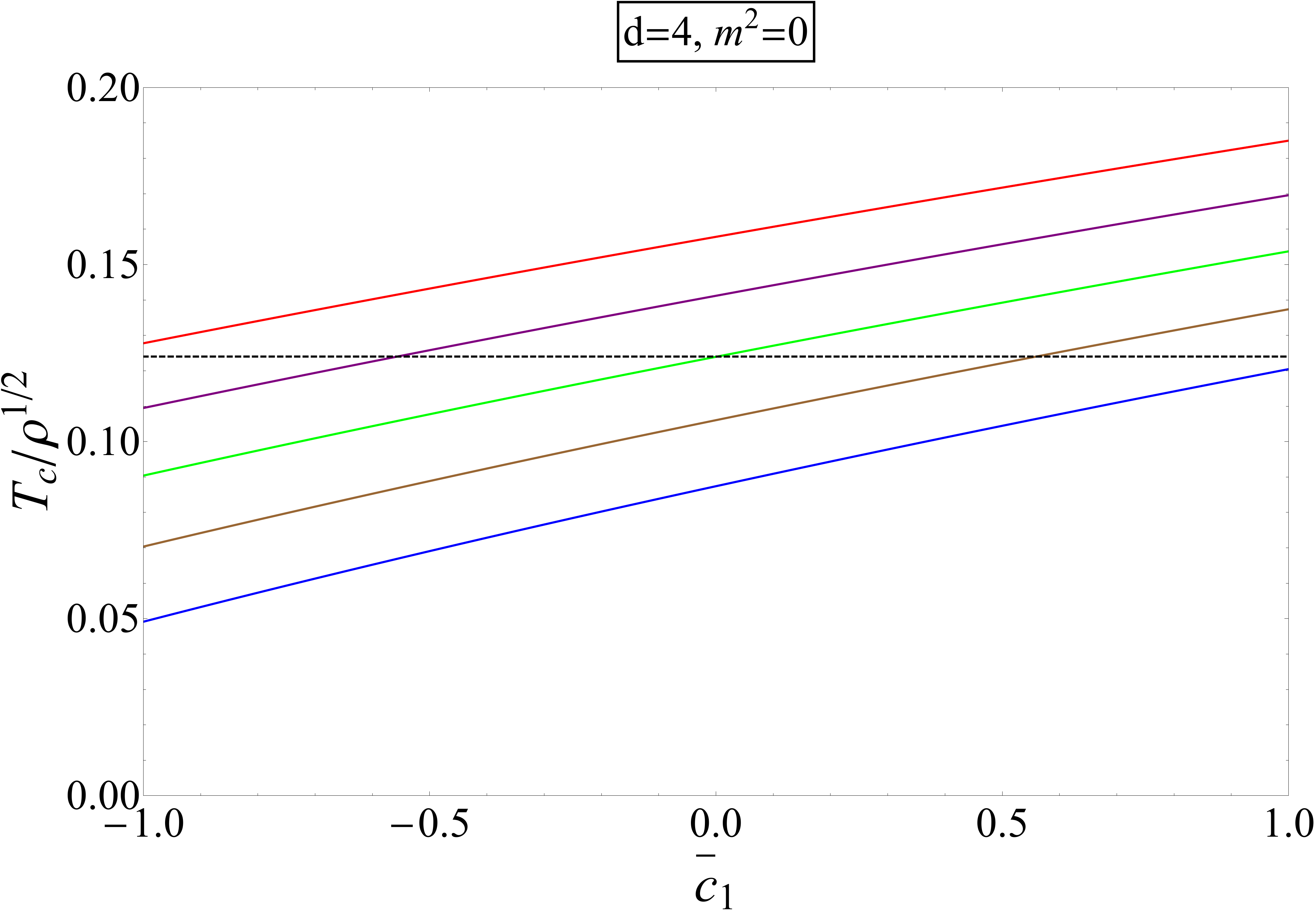}
\hspace*{0.02\textwidth}
\includegraphics[width=0.48 \textwidth]{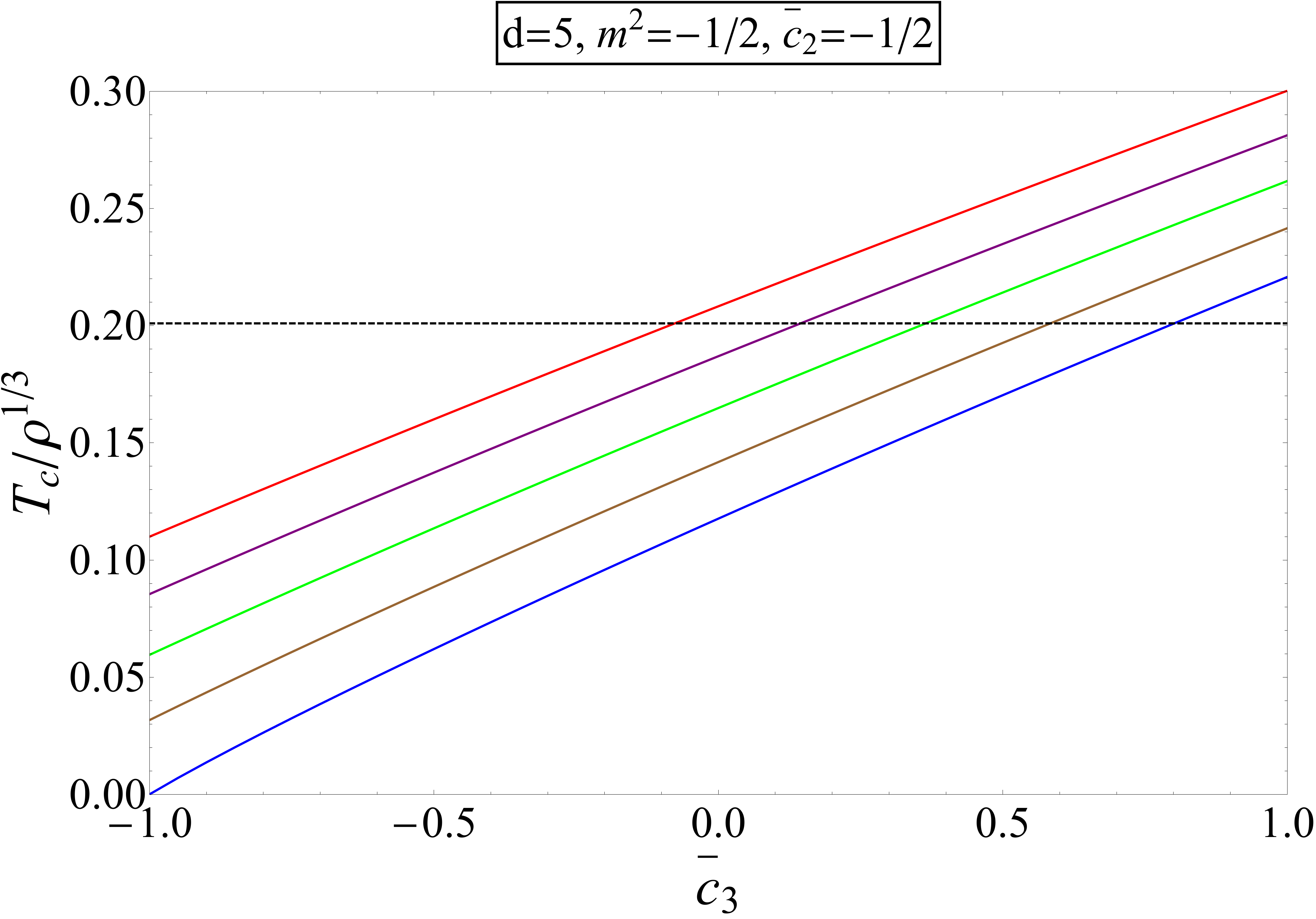}
\end{tabular}
  \caption{The ratio $T_c/\rho^{1/(d-2)}$ as a function of the massive gravity couplings at $m_g=1$. In the left panel: the blue, brown, green, purple, and red curves correspond to $\bar{c}_2$=$-1$, $-0.5$, $0$, $0.5$, and $1$, respectively. In the right panel: the blue, brown, green, purple, and red curves correspond to $\bar{c}_1=-1$, $-0.5$, $0$, $0.5$ and $1$, respectively. The dashed black curves refer to the case of $m_g=0$.}\label{Tc-rho-A}
\end{figure}

In order to understand better the effect of the massive gravity couplings on the critical temperature, we plot the ratio $T_c/\rho^{1/(d-2)}$  as a function of the massive gravity couplings for $d=4$ and $d=5$, as seen in figure \ref{Tc-rho-A}.
We observe that increasing the massive gravity couplings makes the critical temperature increasing. In addition, we see that there exists a critical value for each massive gravity coupling above which the critical temperature in massive gravity is higher than that in Einstein gravity and hence the high temperature superconductors can be achieved in the framework of massive gravity. Whereas, below this critical value the critical temperature in massive gravity is lower than that in Einstein gravity, and hence the $p$-wave superconductor phase transition is harder to achieve in massive gravity. For example, in the left panel of figure \ref{Tc-rho-A}, the critical value of $\bar{c}_1$ corresponding to the purple curve is about $-0.553$ for $\bar{c}_2$ kept fixed. Whereas, in the right panel, for $\bar{c}_{1,2}$ kept fixed, the critical value of $\bar{c}_3$ corresponding to the purple curve is about $0.148$. We note that at such critical values of the massive gravity couplings the term $\bar{c}_1+(d-3)\bar{c}_2+(d-3)(d-4)\bar{c}_3+(d-3)(d-4)(d-5)\bar{c}_4$ is nonzero, but the contribution of massive gravity in the critical temperature via this term would compensate with its contribution via $\lambda_{\text{min}}$, and thus the critical temperature in massive gravity is equal to that in Einstein gravity.

\begin{figure}[t]
% Requires \usepackage{graphicx}\
 \centering
\begin{tabular}{cc}
\includegraphics[width=0.48 \textwidth]{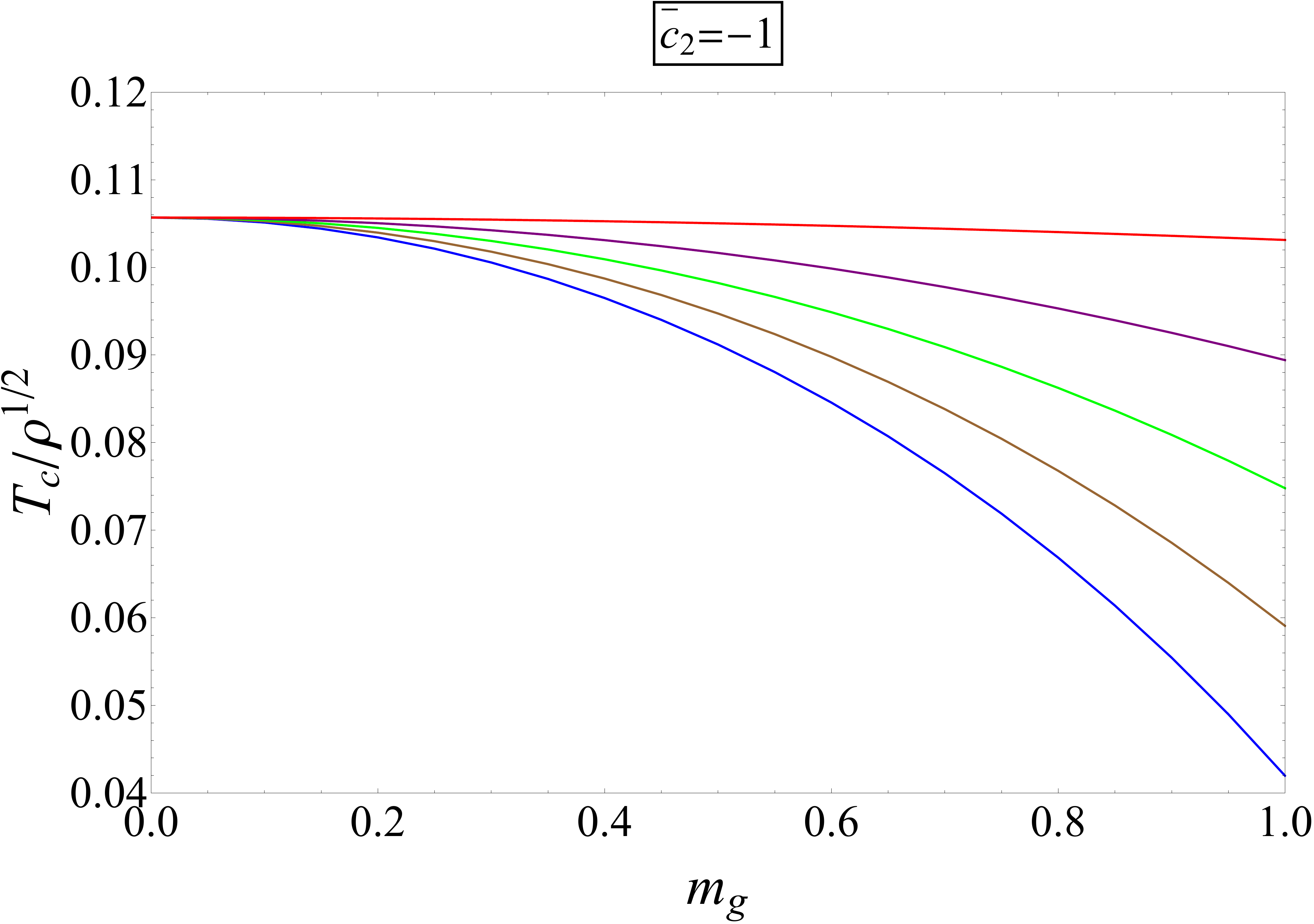}
\hspace*{0.02\textwidth}
\includegraphics[width=0.48 \textwidth]{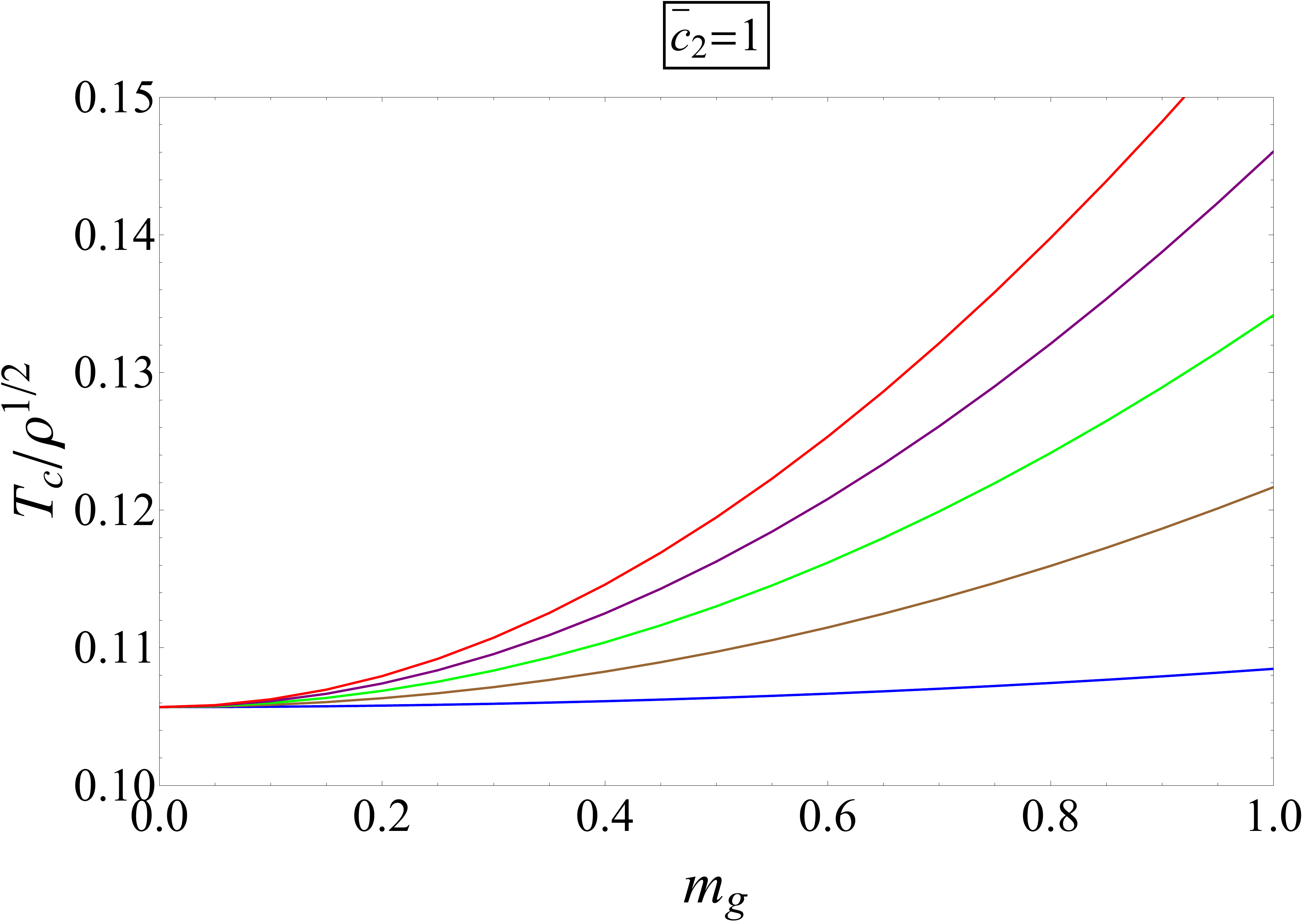}
\end{tabular}
  \caption{The ratio $T_c/\rho^{1/2}$ as a function of the mass of graviton at $d=4$ and $m^2=1/2$. The blue, brown, green, purple, and red curves correspond to $\bar{c}_1=-1$, $-0.5$, $0$, $0.5$, and $1$, respectively.}\label{Tc-rho-B}
\end{figure}

Furthermore, in order to understand better the effect of the mass of graviton on the critical temperature, we plot the critical temperature in terms of the mass of graviton for the remaining parameters kept fixed, given in figure \ref{Tc-rho-B}.
In the case of that the term $\bar{c}_1+(d-3)\bar{c}_2+(d-3)(d-4)\bar{c}_3+(d-3)(d-4)(d-5)\bar{c}_4$ is zero (or small enough), we see from this figure that the red curve in the left panel decreases but the blue curve in the right panel increases with the growth of the mass of graviton. This suggests that in this case the behavior of the critical temperature in massive gravity is either increasing or decreasing function of the mass of graviton depending on that the presence of massive gravity makes $\lambda_{\text{min}}$ decreasing or increasing. However, for the sufficiently large amplitude of the term $\bar{c}_1+(d-3)\bar{c}_2+(d-3)(d-4)\bar{c}_3+(d-3)(d-4)(d-5)\bar{c}_4$, the behavior of the critical temperature depends crucially on the sign of this term. If this term is (negative)positive then the critical temperature (decreases)increases with the growth of the mass of graviton.

In tables \ref{table-Tc-rho-d=4} and \ref{table-Tc-rho-d=5}, we compare the analytical and numerical results which are obtained by using the Sturm-Liouville and numerical methods, respectively.  We see that the analytical results are good agreement with  numerical ones and hence the Sturm-Liouville method is valid to investigate $p$-wave holographic superconductor in massive gravity.
\begin{table}[!htp]
\centering
\begin{tabular}{|c|c|c|c|c|c|}
  \hline
  \multicolumn{3}{|c|}{$\bar{c}_1=1$} & \multicolumn{3}{|c|}{$\bar{c}_2=2$}\\
  \hline
  $ \ \ \bar{c}_2\ \ $ & $ \ \ \text{Analytical} \ \ $ & $\ \ \text{Numerical}\ \ $ & $\ \ \bar{c}_1\ \ $ & $\ \ \text{Analytical}\ \ $ & $\ \ \text{Numerical}\ \ $ \\
\hline
$-1$ & $0.120\rho^{1/2}$ & $0.122\rho^{1/2}$ & $-2$ & $0.132\rho^{1/2}$ & $0.133\rho^{1/2}$\\
\hline
$-0.5$ & $0.137\rho^{1/2}$ & $0.139\rho^{1/2}$ & $-1$ & $0.162\rho^{1/2}$ & $0.165\rho^{1/2}$\\
\hline
$0$ & $0.154\rho^{1/2}$ & $0.156\rho^{1/2}$ & $0$ & $0.190\rho^{1/2}$ & $0.194\rho^{1/2}$\\
\hline
$0.5$ & $0.170\rho^{1/2}$ & $0.173\rho^{1/2}$ & $1$ & $0.215\rho^{1/2}$ & $0.222\rho^{1/2}$\\
\hline
$1$ & $0.185\rho^{1/2}$ & $0.190\rho^{1/2}$ & $2$ & $0.238\rho^{1/2}$ & $0.248\rho^{1/2}$\\
\hline
  \end{tabular}
\caption{Analytical and numerical values for the critical temperature $T_c$ for various values of $\bar{c}_{1,2}$, at $d=4$, $m_g=1$ and $m^2=0$ corresponding to $\Delta_+=1$.}\label{table-Tc-rho-d=4}
\end{table}

\begin{table}[!htp]
\centering
\begin{tabular}{|c|c|c|c|c|c|c|c|c|}
  \hline
  \multicolumn{3}{|c|}{$\bar{c}_2=-\bar{c}_3=1$} & \multicolumn{3}{|c|}{$\bar{c}_3=-\bar{c}_1=1$} & \multicolumn{3}{|c|}{$\bar{c}_1=-\bar{c}_2=1$}\\
  \hline
  $ \ \ \bar{c}_1\ \ $ & $ \ \ \text{Analytical} \ \ $ & $\ \ \text{Numerical}\ \ $ & $\ \ \bar{c}_2\ \ $ & $\ \ \text{Analytical}\ \ $ & $\ \ \text{Numerical}\ \ $ & $\ \ \bar{c}_3\ \ $ & $\ \ \text{Analytical}\ \ $ & $\ \ \text{Numerical}\ \ $ \\
\hline
$-1$ & $0.154\rho^{1/3}$ & $0.155\rho^{1/3}$ & $-1$ & $0.161\rho^{1/3}$ & $0.161\rho^{1/3}$ & $-1$ & $0.055\rho^{1/3}$ & $0.055\rho^{1/3}$\\
\hline
$-0.5$ & $0.175\rho^{1/3}$ & $0.177\rho^{1/3}$ & $-0.5$ & $0.206\rho^{1/3}$ & $0.207\rho^{1/3}$ & $-0.5$ & $0.105\rho^{1/3}$ & $0.105\rho^{1/3}$\\
\hline
$0$ & $0.196\rho^{1/3}$ & $0.197\rho^{1/3}$ & $0$ & $0.250\rho^{1/3}$ & $0.252\rho^{1/3}$ & $0$ & $0.152\rho^{1/3}$ & $0.153\rho^{1/3}$\\
\hline
$0.5$ & $0.215\rho^{1/3}$ & $0.218\rho^{1/3}$ & $0.5$ & $0.290\rho^{1/3}$ & $0.295\rho^{1/3}$ & $0.5$ & $0.198\rho^{1/3}$ & $0.199\rho^{1/3}$\\
\hline
$1$ & $0.234\rho^{1/3}$ & $0.238\rho^{1/3}$ & $1$ & $0.330\rho^{1/3}$ & $0.338\rho^{1/3}$ & $1$ & $0.242\rho^{1/3}$ & $0.244\rho^{1/3}$\\
\hline
  \end{tabular}
\caption{Analytical and numerical values for the critical temperature $T_c$ for various values of $\bar{c}_{1,2,3}$, at $d=5$, $m_g=1$ and $m^2=0$ corresponding to $\Delta_+=2$.}\label{table-Tc-rho-d=5}
\end{table}

\section{\label{CV}Condensate value}

In this section, we obtain an analytical expression for the condensate value $\langle\mathcal{O}_{x+}\rangle$ and investigate the effect of massive gravity on the behavior of $\langle\mathcal{O}_{x+}\rangle$ in terms of the temperature. In order to do this, we need to study the behavior of the gauge field by solving Eq. (\ref{phi-eq-z}). Near the critical temperature $T_c$, the condensate value $\langle\mathcal{O}_{x+}\rangle$ is small. Hence one can expand the function $\phi(z)$ near $T_c$ in terms of $\langle\mathcal{O}_{x+}\rangle$ as
\begin{eqnarray}
\phi(z)=r_+\lambda\xi(z)+r_+\frac{\langle\mathcal{O}_{x+}\rangle^2}{r^{2\Delta_+}_+}\chi(z),\label{expphi}
\end{eqnarray}
where the function $\chi(z)$ satisfies the boundary condition $\chi(1)=\chi'(1)=0$. By substituting this expansion of $\phi(z)$ and the expression of $\rho_x(z)$ given at Eq. (\ref{rho-exp}) into Eq. (\ref{phi-eq-z}), we obtain
\begin{eqnarray}
\chi''(z)+\frac{4-d}{z}\chi'(z)-\frac{2\lambda\left(1-z^{d-3}\right)}{f(z)}z^{2(\Delta_+-1)}F^2(z)=0,\label{chi-Eq}
\end{eqnarray}
which can be rewritten in the following form 
\begin{eqnarray}
\left[\frac{\chi'(z)}{z^{d-4}}\right]'=\frac{2\lambda\left(1-z^{d-3}\right)}{f(z)}z^{2(1+\Delta_+)-d}F^2(z).
\end{eqnarray}
By integrating this equation with the boundary condition $\phi'(1)=0$, we find
\begin{eqnarray}
\chi'(z)&=&z^{d-4}\int^{z}_1\frac{2\lambda\left(1-\widetilde{z}^{d-3}\right)}{f(\widetilde{z})}\widetilde{z}^{2(1+\Delta_+)-d}F^2(\widetilde{z})d\widetilde{z}.\label{Der-psi}
\end{eqnarray}
Expanding of $\phi(z)$ near the AdS boundary, corresponding to Eq. (\ref{expphi}), is given by
\begin{eqnarray}
\phi(z)=\lambda r_+\left(1-z^{d-3}\right)+\frac{\langle\mathcal{O}_+\rangle^2}{r^{2\Delta_+-1}_+}\left[\chi(0)+\chi'(0)z+\frac{\chi''(0)}{2}z^2+\cdots+\frac{\chi^{(d-3)}(0)}{(d-3)!}z^{d-3}+\cdots\right].\label{phi-asybeh-2}
\end{eqnarray}
By comparing the coefficients of $z^{d-3}$ in the right-hand sides of Eqs. (\ref{phi-asybeh}) and (\ref{phi-asybeh-2}), we find
\begin{eqnarray}
\frac{\rho}{r^{d-2}_+}=\lambda-\frac{\langle\mathcal{O}_{x+}\rangle^2}{r^{2\Delta_+}_+}\frac{\chi^{(d-3)}(0)}{(d-3)!}.
\end{eqnarray}
Finally, we obtain the expression for the condensate value $\langle\mathcal{O}_{x+}\rangle$ as a function of the temperature as
\begin{eqnarray}
\langle\mathcal{O}_{x+}\rangle&=&\left(\frac{4\pi}{d-1+a}\right)^{1+\Delta_+}\sqrt{\frac{d-3}{\mathcal{A}}}T^{1+\Delta_+}\sqrt{\left(\frac{T_c}{T}\right)^{d-2}-1},\nonumber\\
&\simeq&\left(\frac{4\pi}{d-1+a}\right)^{1+\Delta_+}\sqrt{\frac{(d-3)(d-2)}{\mathcal{A}}}T^{1+\Delta_+}_c\sqrt{1-\frac{T}{T_c}},\label{conde-value}
\end{eqnarray}
where
\begin{eqnarray}
a&\equiv& m^2_g\left[\bar{c}_1+(d-3)\bar{c}_2+(d-3)(d-4)\bar{c}_3+(d-3)(d-4)(d-5)\bar{c}_4\right],\nonumber\\
\mathcal{A}&\equiv&2\int^1_0z^{4-d+2\Delta_+}(1-z^{d-3})(1-\beta z^2)^2\left\{1-z^{d-1}+m^2_g\left[\frac{\bar{c}_1z}{d-2}(1-z^{d-2})\right.\right.\nonumber\\
&&\left.\left.+\bar{c}_2z^2(1-z^{d-3})+(d-3)\bar{c}_3z^3(1-z^{d-4})+(d-3)(d-4)\bar{c}_4z^4(1-z^{d-5})\right]\right\}^{-1}dz.
\end{eqnarray}
From this expression, we observe that the effect of massive gravity on the condensate value $\langle\mathcal{O}_{x+}\rangle$ is manifested through the terms $a$ and $\beta$. If the term $a$ is (negative)positive and the contribution of massive gravity through this term is dominant compared to its contribution via the term $\beta$, it would make the denominator $d-1+a$ in Eq. (\ref{conde-value}) (decreasing)increasing and thus $\langle\mathcal{O}_{x+}\rangle$ is (larger)smaller than that in Einstein gravity. This suggests that the condensate gets (harder)easier to form in massive gravity.

\begin{figure}[t]
% Requires \usepackage{graphicx}\
 \centering
\begin{tabular}{cc}
\includegraphics[width=0.48 \textwidth]{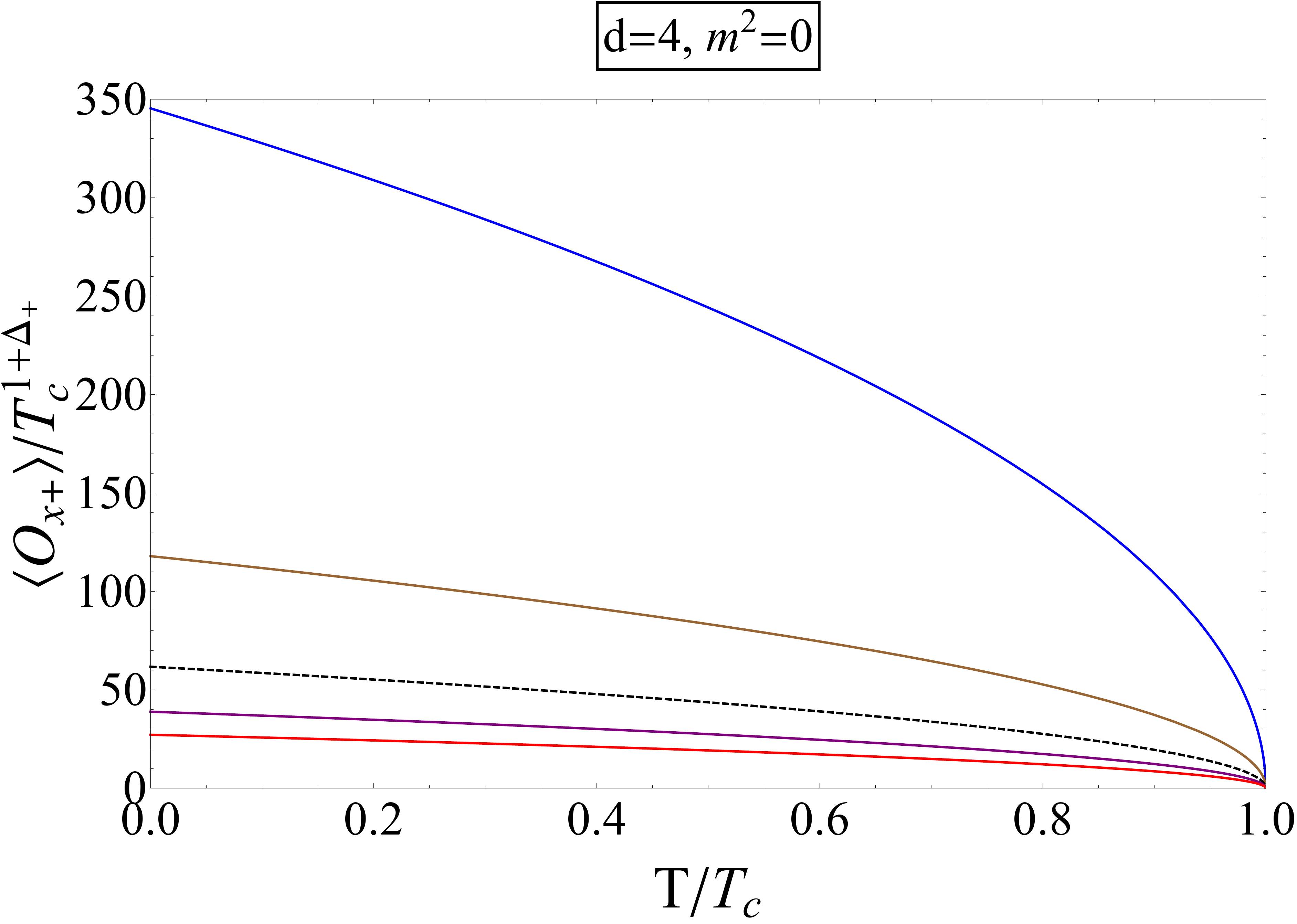}
\hspace*{0.02\textwidth}
\includegraphics[width=0.48 \textwidth]{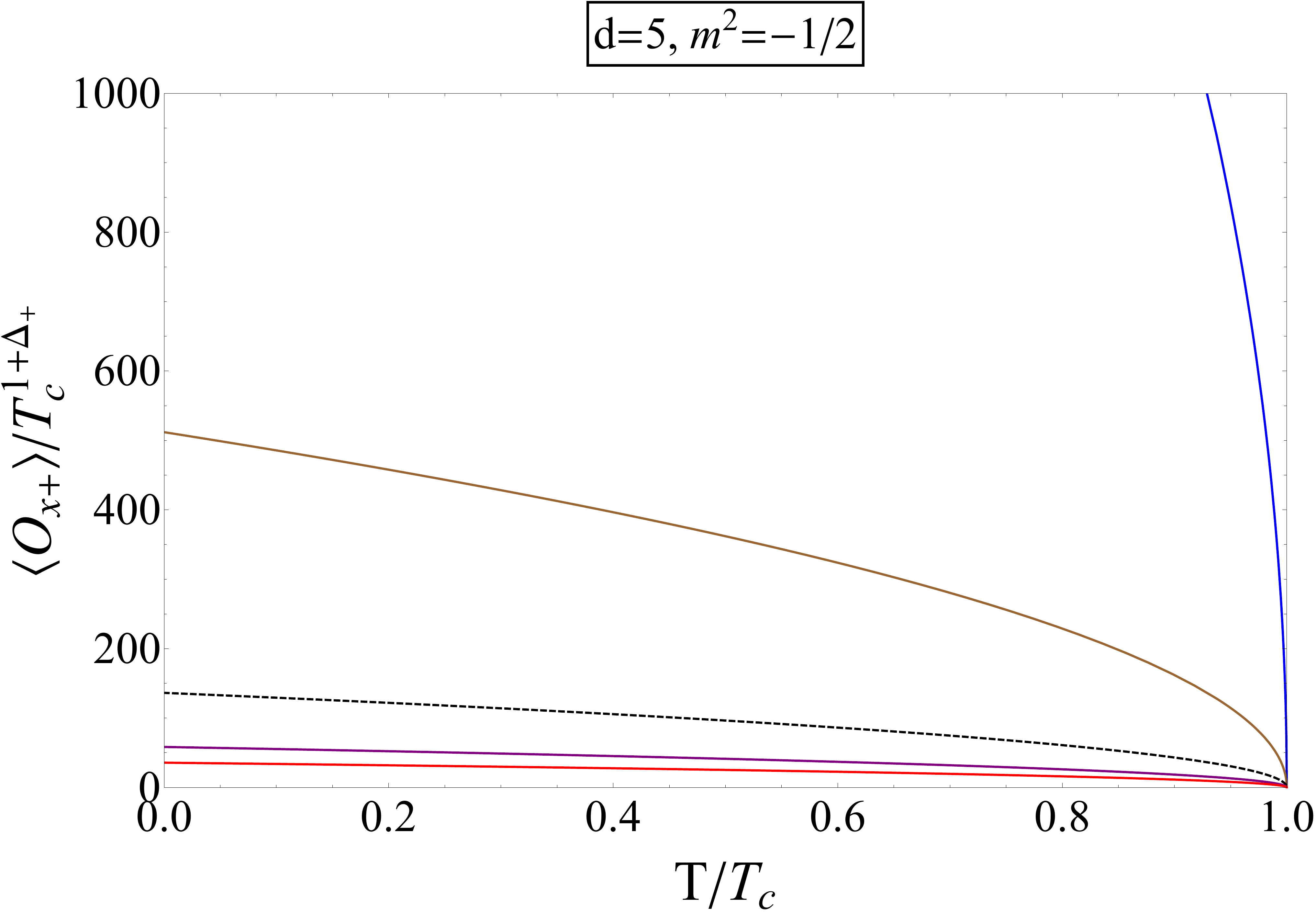}
\end{tabular}
  \caption{The dimensionless condensate value $\langle\mathcal{O}_{x+}\rangle/T^{1+\Delta_+}_c$ as a function of $T/T_c$ at $m_g=1$. In the left panel: the blue, brown, purple, and red curves correspond to the following coupling values $(\bar{c}_1,\bar{c}_2)$=$(-1,-1)$, $(-0.5,-0.5)$, $(0.5,0.5)$, and $(1,1)$, respectively. In the right panel: the blue, brown, purple, and red curves correspond to the following coupling values $(\bar{c}_1,\bar{c}_2,\bar{c}_3)=(-0.5,-0.5,-0.8)$, $(-0.25,-0.25,-0.5)$, $(0.25,0.25,0.5)$, and $(0.5,0.5,0.8)$, respectively. The dashed black curves refer to the case of $m_g=0$.}\label{Ox-T-A}
\end{figure}

\begin{figure}[t]
% Requires \usepackage{graphicx}\
 \centering
\begin{tabular}{cc}
\includegraphics[width=0.48 \textwidth]{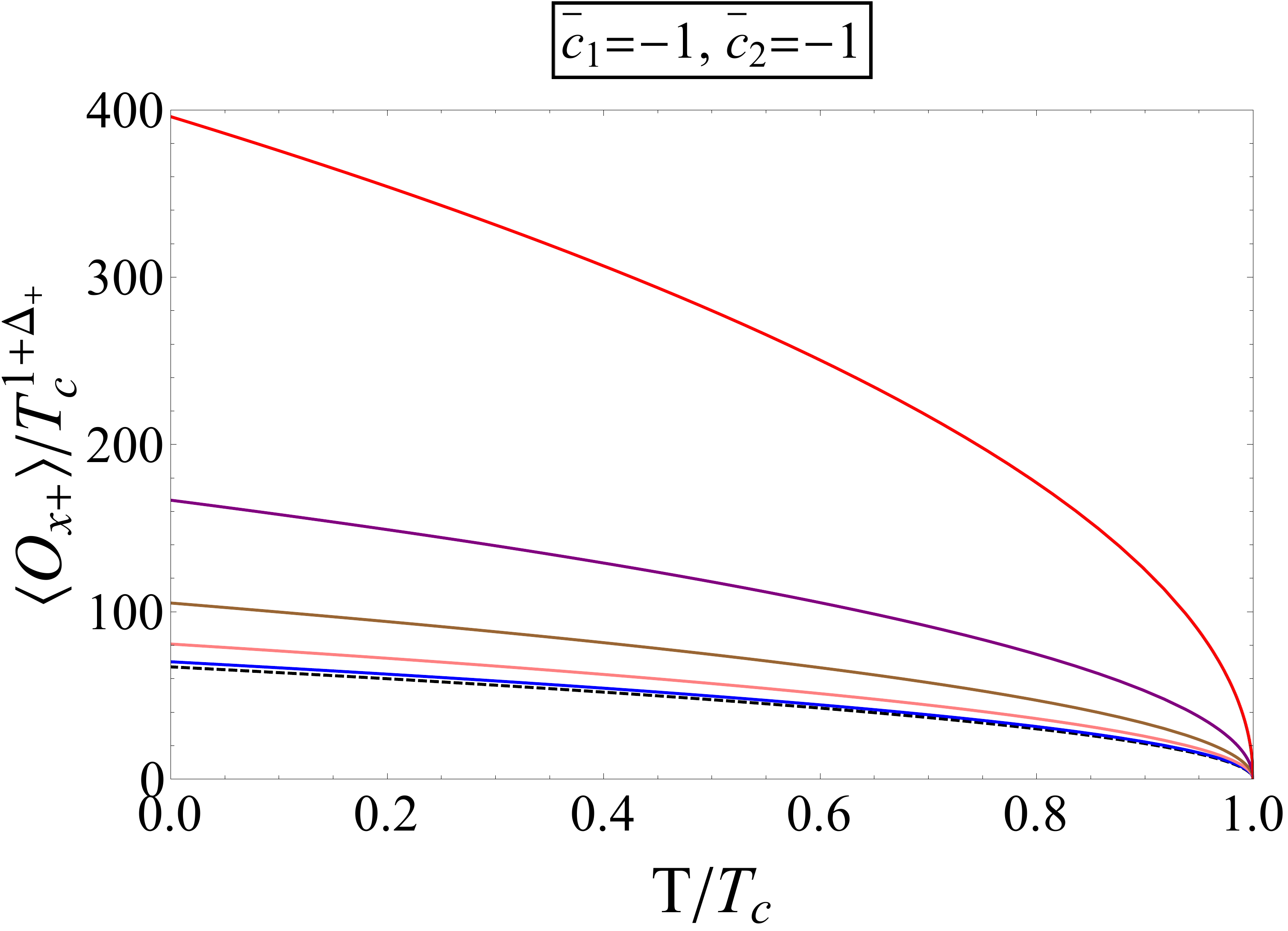}
\hspace*{0.02\textwidth}
\includegraphics[width=0.48 \textwidth]{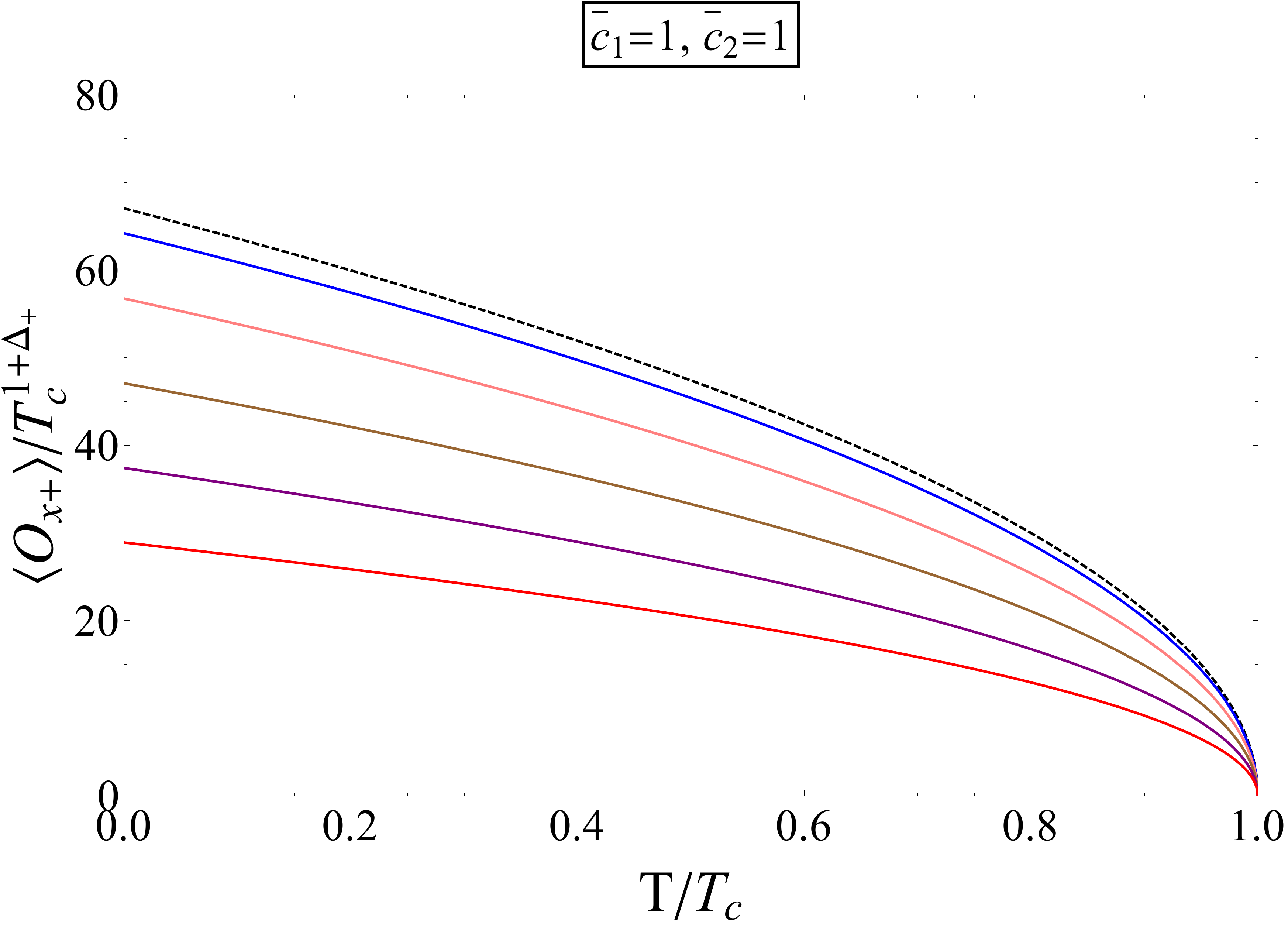}
\end{tabular}
  \caption{The dimensionless condensate value $\langle\mathcal{O}_{x+}\rangle/T^{1+\Delta_+}_c$ as a function of $T/T_c$ at $d=4$ and $m^2=1/2$. The dashed black, blue, pink, brown, purple, and red curves correspond to $m_g=0$, $0.2$, $0.4$, $0.6$, $0.8$, and $1$, respectively.}\label{Ox-T-B}
\end{figure}

In order to see explicitly the effect of the mass of graviton on the condensate value $\langle\mathcal{O}_{x+}\rangle$, we plot the dimensionless condensate value $\langle\mathcal{O}_{x+}\rangle/T^{1+\Delta_+}_c$ as a function of $T/T_c$ for various values of the massive gravity parameters in figure \ref{Ox-T-A} and \ref{Ox-T-B}. We see from figure \ref{Ox-T-A} that, when increasing the massive gravity couplings, the value of the condensate operator becomes smaller. This behavior of the condensate value $\langle\mathcal{O}_{x+}\rangle$ happens similarly if the mass of graviton increases with the sufficiently positive term $\bar{c}_1+(d-3)\bar{c}_2+(d-3)(d-4)\bar{c}_3+(d-3)(d-4)(d-5)\bar{c}_4$, as seen in the right panel of figure \ref{Ox-T-B}. This suggests that the condensate gets easier to form, which is consistent with the behavior of the critical temperature as indicated in the previous section. On the contrary, for decreasing the massive gravity couplings or increasing the mass of graviton with the sufficiently negative term $\bar{c}_1+(d-3)\bar{c}_2+(d-3)(d-4)\bar{c}_3+(d-3)(d-4)(d-5)\bar{c}_4$, the value of the condensate operator becomes larger. On the other hand, in this situation, the condensate gets harder to form.

\section{\label{FE}Free energy for the superconductor and normal phases}
In this section, we study the behavior of the free energy near the critical temperature in the grand canonical ensemble (where the chemical potential is kept fixed) to see the superconductor/normal phase transition. The free energy is given by
\begin{eqnarray}
\Omega=TS_{E},
\end{eqnarray}
where $S_{\text{E}}$ is the Euclidean on-shell action computed as
\begin{eqnarray}
-S_E&=&\frac{i}{2}\int d^dx\sqrt{-g}\left[-\frac{1}{2}\nabla_\mu\left(A_\nu F^{\mu\nu}\right)-\nabla_\mu\left(\rho^\dagger_\nu(D^\mu\rho^\nu-D^\nu\rho^\mu)\right)+\frac{1}{2}A_\nu\nabla_\mu F^{\mu\nu}\right],\nonumber\\
&=&-\frac{V_{d-2}}{2T}\left(-\frac{1}{2}\sqrt{-h}n_rA_\mu F^{r\mu}\Big|_{r\rightarrow\infty}-\sqrt{-h}n_r\rho^\dagger_\mu(D^r\rho^\mu-D^\mu\rho^r)\Big|_{r\rightarrow\infty}\right.\nonumber\\
&&\left.+\frac{1}{2}\int^\infty_{r_+}dr\sqrt{-g}A_\nu\nabla_\mu F^{\mu\nu}\right),\nonumber\\
&=&-\frac{V_{d-2}}{2T}\left[\frac{(d-3)}{2}\mu\rho-\int^{\infty}_{r_+}\frac{\rho^2_x\phi^2}{r^{4-d}f(r)}dr\right],
\end{eqnarray}
where $\int d\tau dx_1...dx_{d-2}=V_{d-2}/T$ with $\tau$ to be the Euclidean time, $h_{ij}=\text{diag}\left[-f(r),r^2,...,r^2\right]$, and $n_\mu=(n_r,n_i)=\left(1/\sqrt{f(r)},\overrightarrow{0}\right)$. Then, the scaled free energy for the superconductor phase is given by
\begin{eqnarray}
\frac{2\Omega_\text{S}}{V_{d-2}}=\frac{d-3}{2}\mu\rho-\int^{\infty}_{r_+}\frac{\rho^2_x\phi^2}{r^{4-d}f(r)}dr.
\end{eqnarray} 
In the normal phase, we have $\rho_x=0$, and hence the corresponding scaled free energy is given by $2\Omega_{\text{N}}/V_{d-2}=(d-3)\mu\rho/2$. By using the solution for $\phi(z)$ and $\rho(x)$ obtained the previous sections, one can find the difference of the scaled free energy between the superconductor and normal phases as
\begin{eqnarray}
\frac{\Delta\Omega}{V_{d-2}}\equiv\frac{2(\Omega_S-\Omega_N)}{V_{d-2}}&\simeq&-\left(\frac{4\pi}{d-1+a}\right)^{d-1}\frac{(d-3)(d-2)}{\mathcal{A}}\left(\int^1_0\frac{\lambda^2z^{4-d+2\Delta_+}\xi^2(z)F^2(z)}{g(z)}dz\right)\nonumber\\
&&\times T^{d-1}_c\left(1-\frac{T}{T_c}\right).
\end{eqnarray}

\begin{figure}[t]
% Requires \usepackage{graphicx}\
 \centering
\begin{tabular}{cc}
\includegraphics[width=0.48 \textwidth]{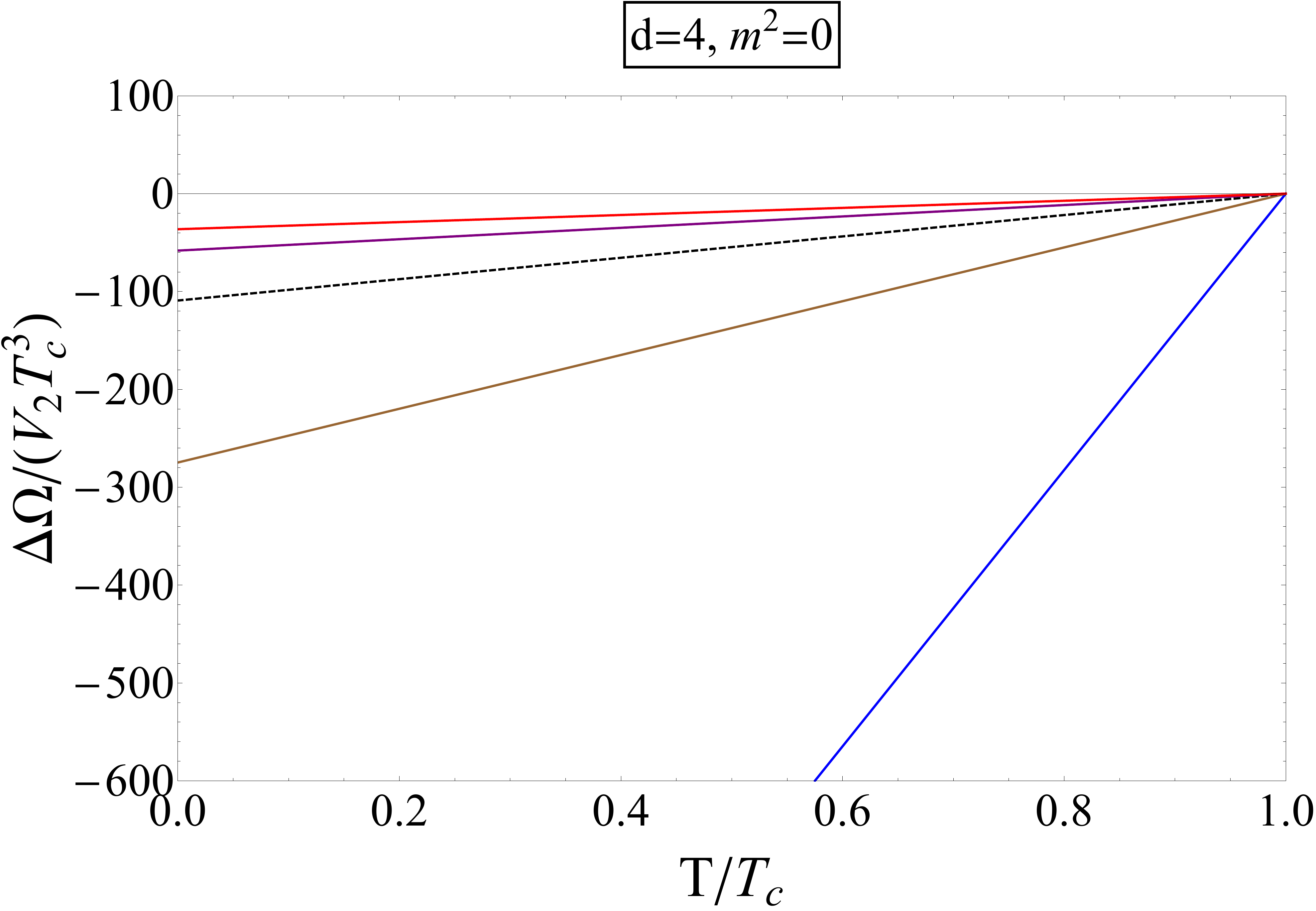}
\hspace*{0.02\textwidth}
\includegraphics[width=0.48 \textwidth]{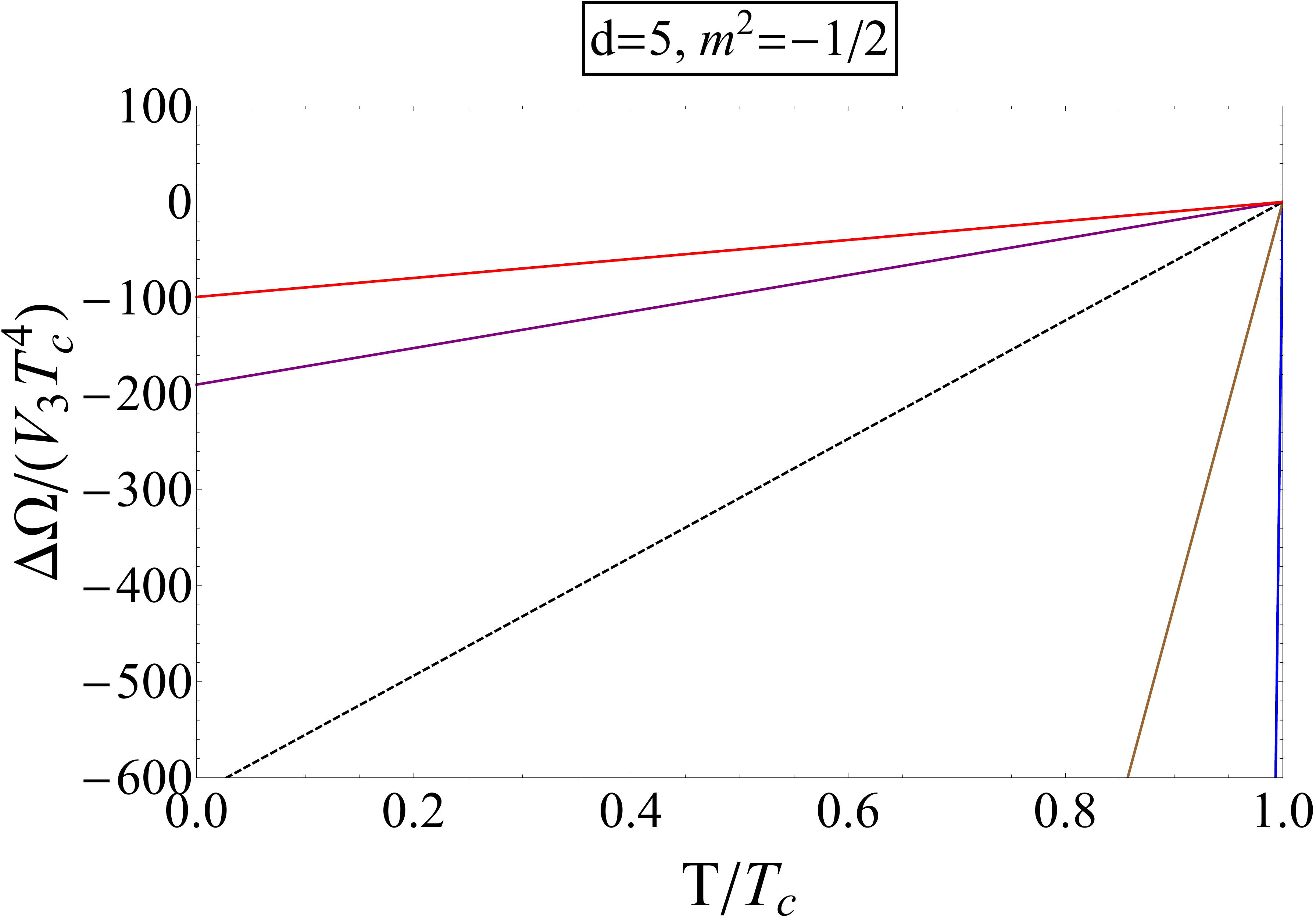}
\end{tabular}
  \caption{The difference of the scaled free energy between the superconductor and normal phases as a function of $T/T_c$ at $m_g=1$. In the left panel: the blue, brown, purple, and red curves correspond to the following coupling values $(\bar{c}_1,\bar{c}_2)$=$(-1,-1)$, $(-0.5,-0.5)$, $(0.5,0.5)$, and $(1,1)$, respectively. In the right panel: the blue, brown, purple, and red curves correspond to the following coupling values $(\bar{c}_1,\bar{c}_2,\bar{c}_3)=(-0.5,-0.5,-0.8)$, $(-0.25,-0.25,-0.5)$, $(0.25,0.25,0.5)$, and $(0.5,0.5,0.8)$, respectively. The dashed black curves refer to the case of $m_g=0$.}\label{free-en-A}
\end{figure}

\begin{figure}[t]
% Requires \usepackage{graphicx}\
 \centering
\begin{tabular}{cc}
\includegraphics[width=0.48 \textwidth]{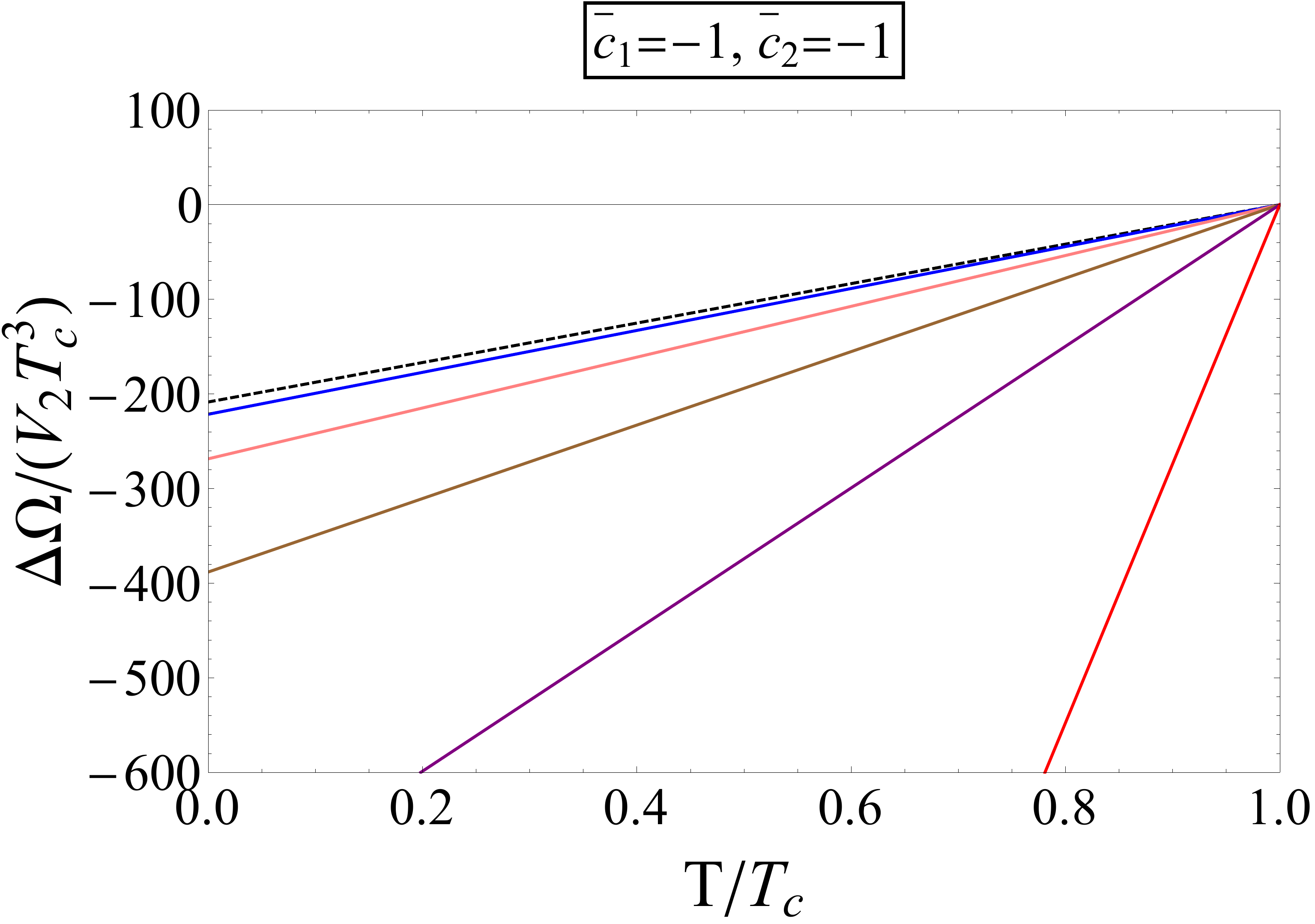}
\hspace*{0.02\textwidth}
\includegraphics[width=0.48 \textwidth]{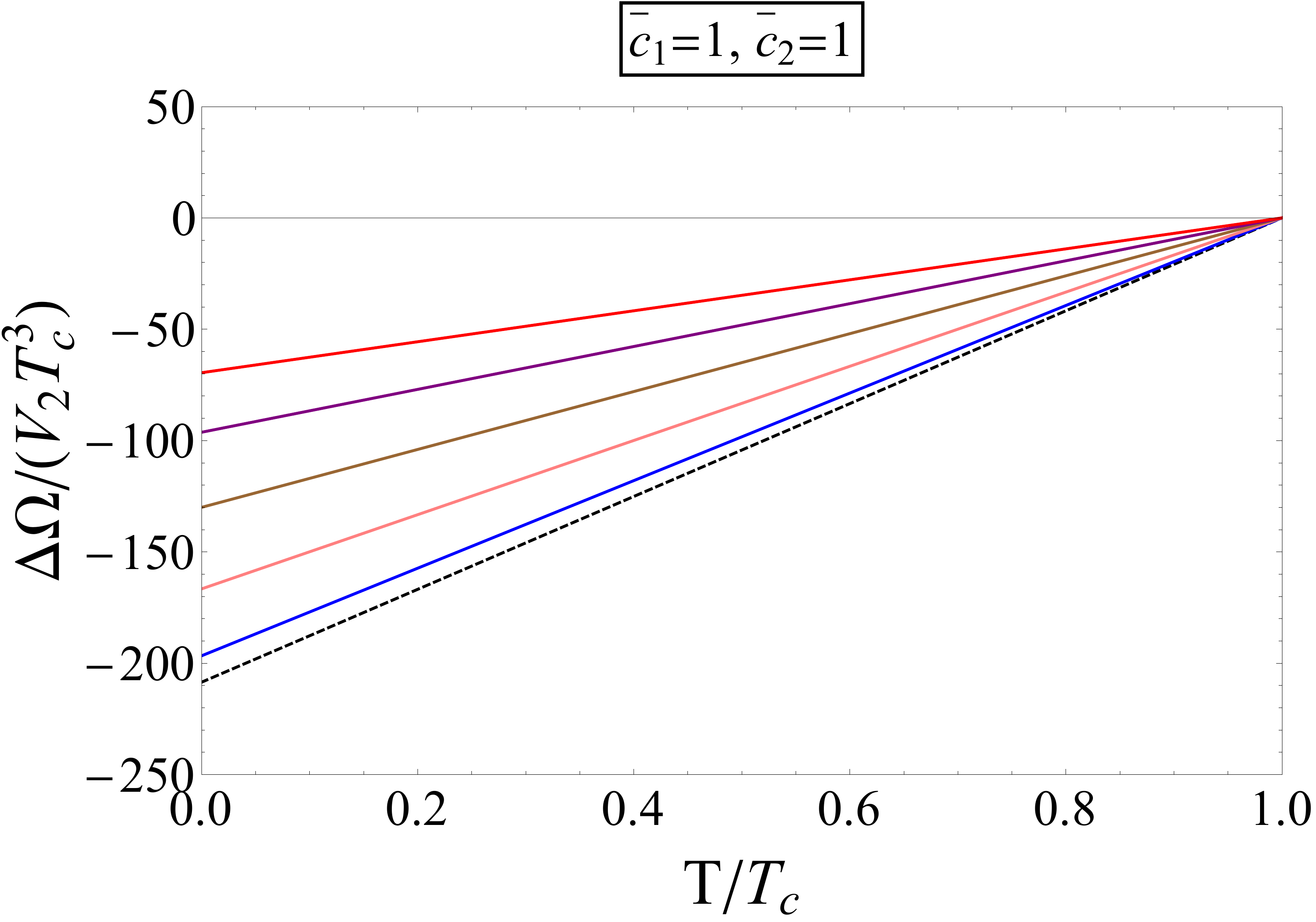}
\end{tabular}
  \caption{The difference of the scaled free energy between the superconductor and normal phases as a function of $T/T_c$ at $d=4$ and $m^2=1/2$. The dashed black, blue, pink, brown, purple, and red curves correspond to $m_g=0$, $0.2$, $0.4$, $0.6$, $0.8$, and $1$, respectively.}\label{free-en-B}
\end{figure}

We show the difference of the scaled free energy between the superconductor and normal phases in terms of $T/T_c$ in the figures \ref{free-en-A} and \ref{free-en-B}. These figures indicate that the superconductor phase has the free energy smaller than the normal phase and hence it contributes dominantly to the thermodynamics. In this way, below the critical temperature the superconductor phase rather than normal phase is thermodynamically favored. In addition, decreasing the massive gravity couplings or increasing the mass of graviton with the sufficiently negative term $\bar{c}_1+(d-3)\bar{c}_2+(d-3)(d-4)\bar{c}_3+(d-3)(d-4)(d-5)\bar{c}_4$ leads to that the superconductor phase is more stable. This is because as indicated in the previous section the value of the condensate operator in this case gets larger with the decreasing of the massive gravity couplings or the growth of the mass of graviton, and thus much more energy is needed to break the condensate.

\section{\label{conclu} Conclusion}
In this paper, we have analytically studied the effects of massive gravity on $p$-wave holographic superconductor in the probe limit which the backreaction of the matter fields on the spacetime geometry is ignored. We found that the massive gravity parameters affect significantly on the critical temperature and the value of the order parameter in order:
\begin{itemize}

\item When increasing(decreasing) the massive gravity couplings, the critical temperature increases(decreases), whereas the value of the order parameter gets lower(higher).

\item Above a critical value of each massive gravity coupling, the critical temperature in massive gravity is higher than that in Einstein gravity, whereas the value of the order parameter is smaller compared to the case of massless graviton.

\end{itemize}
These facts suggest that the high temperature superconductors can be achieved in the presence of graviton mass with the proper parameters. In addition, we indicated that for the negative massive gravity couplings, the superconductor phase with the stronger coupling amplitude or the more massive graviton is more thermodynamically favored.

Note added.-- When this paper was being finalized, a work recently appeared \cite{NieZeng2020} which studied the $p$-wave holographic superfluid model in massive gravity, using the numerical method.

\end{document}